# LIMIT THEOREMS FOR QUANTUM WALKS ON THE UNION OF PLANES

**CLEMENT AMPADU**

31 Carrolton Road
Boston, Massachusetts, 02132
U.S.A.
e-mail: drampadu@hotmail.com

## Abstract

We extend the construction given by [Chisaki et.al, arXiv:1009.1306v1] from lines to planes, and obtain the associated limit theorems for quantum walks on such a graph.

## I. Introduction

This paper adds to the growing literature on limit theorems for quantum walks. These theorems seek to explain observations in numerical simulations of quantum random walks whereby the probability distribution of the walkers position is seen to exhibit a persistent major spike at the initial position and two other minor spikes which drift to infinity in either direction. Under the pseudonym "localization" theoretical explanations for this phenomenon has been given by many authors in various contexts, see [1-22] for examples. In this paper we consider a discrete time quantum walk $P_{t,k}$ at time $t$ on a graph with joined quarter planes $P_t$, which is composed of $k$ quarter planes with the same origin. Employing the reduction technique of Chisaki et.al [5], we obtain two types of limit theorems. The first is an asymptotic behavior of $P_{t,k}$ which corresponds to localization (Theorem 1). The second is a weak convergence statement for $P_{t,k}$ (Theorem 2).

This paper is organized as follows. In Sections II, III, IV, and V we give the definitions of the discrete-time quantum walk on the plane, the quarter plane, a graph with joined quarter planes, and a new type of homogeneous tree. Noting that the plane is a regular graph, in the introduction of Section III we express a general definition of the discrete-time quantum walk on the undirected graph. Sections V and VI is mainly concerned with introducing a quantum walk with an enlarged basis and the reduction of $P_{t,k}$ to the walk on a quarter plane. Lemma 1 gives the reduction of $T_{t,k_x,k_y}$ to $P_{t,k}$ whilst Lemma 2 extends the state of $P_{t,k}$ to $P'_{t,k}$. In Section VII we obtain the generating function of the walk, and use it in Section VIII to prove Theorem 1. In Section IX we prove Theorem 2 using the Fourier transform of the generating function. Section X is devoted to the conclusions.

## II. Discrete-time quantum walk on the plane

Let $Z$ be the set of integers, and let $Z \times Z = \{(x, y) : x, y \in Z\}$. The quantum walk on $Z \times Z$ is defined by $H = H_P^{Z \times Z} \otimes H_C^{Z \times Z}$, where $H_P^{Z \times Z} = span\{|x, y\rangle : x, y \in Z\}$ and

$H_C^{Z \times Z} = span\{|Left\rangle, |Right\rangle, |Up\rangle, |Down\rangle\}$. Put $|Left\rangle =^T [1 \ 0 \ 0 \ 0]$,

$|Right\rangle =^T [0 \ 1 \ 0 \ 0]$, $|Down\rangle =^T [0 \ 0 \ 1 \ 0]$, $|Up\rangle =^T [0 \ 0 \ 0 \ 1]$, where $T$ is the transposed operator. Recall part of the unitary operator is given by $F_{Z \times Z} = I_P \otimes C_{Z \times Z}$ where $I_P$ is the identity operator on $H_P^{Z \times Z}$, and $C_{Z \times Z} = \begin{bmatrix} a^2 & ab & ab & b^2 \\ ac & ad & bc & bd \\ ac & bc & ad & bd \\ c^2 & cd & cd & d^2 \end{bmatrix} \in U(4)$ with $abcd \neq 0$,

where $U(4)$ is the set of $4 \times 4$ unitary matrices. The shift operator $S_{Z \times Z}$ is defined by

$$S_{Z \times Z}|x, y, l\rangle = \begin{cases} |x-1, y, left\rangle, l = left \\ |x+1, y, right\rangle, l = right \\ |x, y-1, down\rangle, l = down \\ |x, y+1, up\rangle, l = up \end{cases}$$

## III. Discrete-time quantum walk on the quarter plane

Note that the plane is a regular graph. In general given an undirected connected graph $G$. Let $V(G)$ be the set of all sites in $G$. Let $E_{x,y}(G) \subset E(G)$ be the set of all bonds which connect the site $(x, y) \in V(G)$. Let the position subpace be given by $H_P = span\{|x, y\rangle : (x, y) \in V(G)\}$. For $(x, y) \in V(G)$ let the coin subspace be given by $H_{C_{(x,y)}} = span\{|l\rangle : l \in E_{x,y}(G)\}$. A discrete-time quantum walk on $G$ is given by
$H = span\{|x, y, l\rangle : (x, y) \in V(G) \text{ and } l \in E_{(x,y)}(G)\} = H_P \otimes H_{C_{(x,y)}}$. If $G$ is a regular graph, then $H = H_P \otimes H_{C_{(0,0)}}$. In general for the quantum walk on the undirected connected graph $G$, the evolution operator on the space $H = H_P \otimes H_{C_{(x,y)}}$ is determined by $U = SF$, where $S : H \to H$ is a shift operator and $F : H \to H$ is defined by $F = \sum_{(x,y) \in V(G)} |x, y\rangle\langle x.y| \otimes C_{(x,y)}$, where $C_{(x,y)}$ is the coin operator on $H_{C_{(x,y)}}$. If $G$ is a regular graph and for every $(x, y) \in V(G)$, $C_{(x,y)}$ is independent of $(x, y)$, then we can write $F = I_P \otimes C$, where $I_P$ is the identity operator on $H_P$. In general for the quantum walk on the undirected connected graph $G$, $U = SF$ is a unitary operator if and only if $S$ is a permutation on $H = H_P \otimes H_{C_{(x,y)}}$ and all $C_{(x,y)}$ are unitary operators. We should remark that the state of the walker is given by $\Psi_t(x, y) = \sum_{l \in E_{(x,y)}(G)} \alpha_t(x, y, l) |x, y, l\rangle$, where $\alpha_t(x, y, l) \in C$ is the amplitude of the base $|x, y, l\rangle$ at time $t$ and $C$ is the set of complex numbers. The probability of finding the walker at $(x, y)$ at time $t$ is given by $\|\Psi_t(x, y)\|^2 = \sum_{l \in E_{x,y}(G)} |\alpha_t(x, y, l)|^2$. Let us consider initial states starting from the origin, in particular let $\Psi_0(0,0)$ be such that $\|\Psi_0(0,0)\| = 1$.

Concerning the quarter plane which we denote by $Q$, it is a graph with sites $V(Q) = \{0,0\} \cup \{(x, y) : x \in Z_+, y \in Z_+, Z_+ = \{1, 2, \cdots\}\}$. Note that $\{(x, y) : x \in Z_+, y \in Z_+, Z_+ = \{1, 2, \cdots\}\}$ is the Cartesian product $Z_+ \times Z_+$. We say sites $(x_1, y_1)$ and $(x_2, y_2)$ connect if and only if $\sqrt{(x_1 - x_2)^2 + (y_1 - y_2)^2} = 1$. Let $Q_t$ be the quantum walk on $Q$ whose Hilbert space $H^Q = span\{|x, y, l\rangle : x, y \in Z_+, l \in \{Right, Left, Up, Down\}\} \cup |0,0, \in\rangle\}$. As mentioned earlier, we put $|Left\rangle =^T [1\ 0\ 0\ 0]$, $|Right\rangle =^T [0\ 1\ 0\ 0]$, $|Down\rangle =^T [0\ 0\ 1\ 0]$, $|Up\rangle =^T [0\ 0\ 0\ 1]$, where $T$ is the transposed operator. We should note that the unitary operator is given by $U = S_Q F_Q$, where $F_Q = |0,0\rangle\langle 0,0| \otimes \tilde{c} + \sum_{x,y} |x, y\rangle\langle x, y| \otimes C_{Z \times Z}$, where $\tilde{c} \in C$ is such that $|\tilde{c}| = 1$ and $C_{Z \times Z}$ is as given earlier on. The shift operator $S_Q$ is given by $S_Q|0,0, \in\rangle = |1,0, Right\rangle + |0,1, Up\rangle$,

$$S_Q|1,0,l\rangle = \begin{cases} |0,0,\in\rangle, l = left \\ |2,0, Right\rangle, l = Right \end{cases}, \quad S_Q|0,1,l\rangle = \begin{cases} |0,0,\in\rangle, l = Down \\ |0,2, Up\rangle, l = Up \end{cases},$$

$$S_Q|x, y, l\rangle = \begin{cases} |x-1, y, left\rangle, l = Left \\ |x+1, y, Right\rangle, l = Right \\ |x, y-1, Down\rangle, l = Down \\ |x, y+1, Up\rangle, l = Up \end{cases}, x \geq 2, y \geq 2$$

## IV. Discrete-time quantum walk on the "join" of quarter planes

Let the graph with joined quarter planes be denoted by $P_k$ and let $P_{t,k}$ be the quantum walk on $P_k$. Define $K_k = \{0,1,2,\cdots,k-1\}$. For $r \in K_k$ with $k \geq 1$, put $k$ – quarter planes $Q_r$ by $V(Q_r) = \{h_r(0,0)\} \cup (Z_+ \times Z_+)_r$, where $(Z_+ \times Z_+)_r = \{h_r(x,y) : x \in Z_+, y \in Z_+, Z_+ = \{1,2,\cdots\}\}$. We say sites $h_r(x_1, y_1)$ and $h_r(x_2, y_2)$ connect if and only if $\sqrt{(x_1-x_2)^2 + (y_1-y_2)^2} = 1$. We define $P_k$ as $V(P_k) = \bigcup_j V(Q_j)$, and $E(P_k) = \bigcup_j E(Q_j)$ with $h_r(0,0) \equiv 0$ for any $r$. Let $(0,0)$ be the origin of $P_k$ so that the quarter planes join at $(0,0)$. Let

$$H^{P_k} = span\{\{|x,y,l\rangle : (x,y) \in V(P_k) \setminus \{(0,0)\}, l \in \{Up, Down, Right, Left\}\} \cup \{|0,0,l\rangle : l \in \{\in_0, \cdots, \in_{k-1}\}\}\}$$

be Hilbert space of the quantum walk on $P_k$, where $\in_r = \begin{bmatrix} \overbrace{0\ldots0}^{r} & 1 & \overbrace{0\ldots0}^{k-r-1} \end{bmatrix}^T$. The unitary operator is given by $U = S_P F_P$; where $F_P = |0,0\rangle\langle 0,0| \otimes \tilde{c} G_k + \sum_{x,y \in V(P_k)/\{(0,0)\}} |x,y\rangle\langle x,y| \otimes C_{Z \times Z}$, where $G_k$ will be taken to be the Grover operator, $\tilde{c} \in C$, with $|\tilde{c}| = 1$, and $C_{Z \times Z}$ is as defined earlier. The shift operator $S_P$ is given by $S_P|0,0,\in_r\rangle = |h_r(1,0), Right\rangle + |h_r(0,1), Up\rangle$,

$$S_P|h_r(1,0),l\rangle = \begin{cases} |0,0,\in_r\rangle, l = Left \\ |h_r(2,0), Right\rangle, l = Right \end{cases}, \quad S_P|h_r(0,1),l\rangle = \begin{cases} |0,0,\in_r\rangle, l = Down \\ |h_r(0,2), Up\rangle, l = Up \end{cases},$$

$$S_P|h_r(x,y),l\rangle = \begin{cases} |h_r(x-1,y), left\rangle, l = left \\ |h_r(x+1,y), right\rangle, l = right \\ |h_r(x,y-1), down\rangle, l = down \\ |h_r(x,y+1), up\rangle, l = up \end{cases}, x \geq 2, y \geq 2$$

## V. Discrete-time quantum walk on homogeneous trees

Recall the goal of this paper is to obtain the associated limit theorems for quantum walks on the structure defined in the previous section. To adapt the reduction technique of Chisaki et.al [5], we introduce a new type of homogeneous tree. We define a new type of homogeneous tree $V_{k_{x,y}}$, and a quantum walk $T_{t,k_{x,y}}$ on $V_{k_{x,y}}$. Fix $k \geq 2$, and let $\sum_x = \{\sigma_{x_0}, \cdots \sigma_{x_{(k-1)}}\}$, and let $\sum_y = \{\sigma_{y_0}, \cdots \sigma_{y_{(k-1)}}\}$ be the set of generators subject to $\sigma_{x_j}^2 = e_x$ and $\sigma_{y_j}^2 = e_y$ for $j \in K_k$, where $K_k$ is as defined earlier on, and $e_x$ and $e_y$ are the identity of the groups respectively. Let

$V(T_{k_x}) = \{\sigma_{x_{i_n}} \ldots \sigma_{x_{i_1}} : n \geq 0, \sigma_{x_{i_j}} \in \sum_x \text{ and } i_{j+1} \neq i_j \text{ for } j = 1,2,\ldots,n-1\}$,

$V(T_{k_y}) = \{\sigma_{y_{i_n}} \ldots \sigma_{y_{i_1}} : n \geq 0, \sigma_{y_{i_j}} \in \sum_x \text{ and } i_{j+1} \neq i_j \text{ for } j = 1,2,\ldots,n-1\}$, and the Cartesian

product $V(T_{k_x}) \times V(T_{k_y})$ be the set of all sites on $V_{k_{x,y}}$. We will say that sites $g$ and $h$ connect if and only if $gh^{-1} \in \sum_x \times \sum_y$. Let $H_P^{V_{k_{x,y}}} = \text{span}\{|x,y\rangle : (x,y) \in V(T_{k_x}) \times V(T_{k_y})\}$ be the Hilbert space of the position subspace, and let $H_C^{V_{k_{x,y}}} = \text{span}\{|\sigma_{x_j}, \sigma_{y_j}\rangle : \sigma_{x_j} \in \sum_x, \sigma_{y_j} \in \sum_y\}$ be the Hilbert space of the coin subspace. The unitary operator is defined by $U = S_V F_V$ where

$$F_V = |e_x, e_y\rangle\langle e_x, e_y| \otimes \tilde{c} G_k + \sum_{(x,y)\in V(T_{k_x})\times V(T_{k_y})\setminus\{(e_x,e_y)\}} |x,y\rangle\langle x,y| \otimes G_k, \text{ where } \tilde{c} \in C \text{ is such that } |\tilde{c}| = 1,$$

and $S_V|x, \sigma_x, y, \sigma_y\rangle = |\sigma_x x, \sigma_x, \sigma_y y, \sigma_y\rangle$, where $|\sigma_{x_r}\rangle$ or $|\sigma_{y_r}\rangle$ is given by

$$[\overbrace{0...0}^{r} \; 1 \; \overbrace{0...0}^{k-r-1}]^T.$$ We see immediately that we have the following lemma which requires no proof.

**Lemma 1 (Reduction of $T_{t,k_{x,y}}$ to $P_{t,k}$):** The quantum walk on $V_{k_{x,y}}$ is reduced to $P_{t,k}$ with coin operator $F_J^T = |0,0\rangle\langle 0,0| \otimes \tilde{c} G_k + \sum_{(x,y)\in V(P_k)\setminus\{(0,0)\}} |x,y\rangle\langle x,y| \otimes C_k$, where

$$C_k = \begin{bmatrix} (k-1)b_k^2 & a_k b_k \sqrt{k-1} & a_k b_k \sqrt{k-1} & a_k^2 \\ -a_k b_k \sqrt{k-1} & (k-1)b_k^2 & -a_k^2 & a_k b_k \sqrt{k-1} \\ -a_k b_k \sqrt{k-1} & -a_k^2 & (k-1)b_k^2 & a_k b_k \sqrt{k-1} \\ a_k^2 & -a_k b_k \sqrt{k-1} & -a_k b_k \sqrt{k-1} & (k-1)b_k^2 \end{bmatrix}$$

## VI. The Reduction Technique

In order to give the limit theorems for $P_{t,k}$ we consider a reduction of $P_{t,k}$ on the quarter plane. We first introduce $P'_{t,k}$ which is an enlarged basis of $P_{t,k}$. Let $(X_t^*, Y_t^*)$ be associated with the event "$X_t^* = x, Y_t^* = y$", we construct $(X_t^*, Y_t^*)$ as a reduction of $P'_{t,k}$ on a quarter plane. To analyze $(X_t^*, Y_t^*)$ we give the generating function of the states. By using it we shall obtain the limit states and the characteristic function of $P_{t,k}$.

Let $\Psi_t(x,y)$ be the state of the quantum walk $P_{t,k}$ at time $t$ and position $(x,y)$. Define the initial state $\psi$ by $\Psi_0(0,0) = \sum_{j \in K_k} \psi_j |0,0,e_j\rangle$ with $\|\Psi_0(0,0)\| = 1$. Now we consider the state at first step.

For any $r \in K_k$, $\Psi_1(h_r(1,0)) = \sum_{j \in K_k} \tilde{c}\left(\frac{2}{k} - \delta_r(j)\right) \psi_j |h_r(1,0), Right\rangle$ and

$\Psi_1(h_r(0,1)) = \sum_{j \in K_k} \tilde{c}\left(\frac{2}{k} - \delta_r(j)\right) \psi_j |h_r(1,0), Up\rangle$. Let $\psi'_r = \sum_{j \in K_k} \tilde{c}\left(\frac{2}{k} - \delta_r(j)\right) \psi_j$, then we can

write $\Psi_1(h_r(1,0)) = \psi'_r|h_r(1,0), Right\rangle$, and $\Psi_1(h_r(0,1)) = \psi'_r|h_r(0,1), Up\rangle$. Using the orthonormal basis $\{|\in'_j\rangle : j \in K_k\}$, we can write

$$\Psi_1(h_r(1,0)) = \left(\sum_{j \in K_k} \psi'_j\langle\in'_j| \otimes I_P\right)|e'_r\rangle|h_r(1,0), Right\rangle = \Lambda(\psi)|\in'_r\rangle|h_r(1,0), Right\rangle,$$ where $I_P$ is the

identity operator in $H^{P_k}$ and we have let $\Lambda(\psi) = \sum_{j \in K_k} \psi'_j\langle\in'_j| \otimes I_P$. Similarly we can write

$\Psi_1(h_r(0,1)) = \Lambda(\psi)|\in'_r\rangle|h_r(0,1), Up\rangle$. Now let $H'$ be a Hilbert space spanned by an by an orthonormal basis $\{|\in'_j\rangle : j \in K_k\}$, then we can define $P'_{t,k}$ as a quantum walk on $H' \otimes H^{P_k}$ starting at $t = 1$ with the evolution operator $U'_P = I_k \otimes S_P F_P$ and the initial state Is either $\sum_{j \in K_k}|\in'_j\rangle|h_j(1,0), Right\rangle$ or $\sum_{j \in K_k}|\in'_j\rangle|h_j(0,1), Up\rangle$ where $I_k$ is the identity operator on $H'$.

Let $l_0 = \{\in_0, \cdots, \in_{k-1}\}$ and let $l_{x,y} = \{Left, Right, Up, Down\}$ for $(x, y) \in V(P_k) \setminus \{(0,0)\}$. Then the state of the quantum walk $P'_{t,k}$ at time $t$ and position $(x, y)$ is written as

$\Psi'_t(x, y) = \sum_{j \in K_k, u \in U_{x,y}} \alpha'_t(\in'_j, x, y, u)|\in'_j\rangle|x, y, u\rangle$, where $\alpha'_t(a, b, c, d)$ is the amplitude of the base

$|a, b, c, d\rangle$ at time $t$. In particular we have the following.

**Lemma 2 (Extending the state of $P_{t,k}$ to $P'_{t,k}$):** For any $t \geq 1$ and $(x, y) \in V(P_k)$,

$\Psi_t(x, y) = \Lambda(\psi)\Psi'_t(x, y)$.

**Proof:** We show by induction on $t$. If $t = 1$, there is nothing to prove. Fix $t \geq 2$, and assume $\Psi_t(x, y) = \Lambda(\psi)\Psi'_t(x, y)$, then for any $(x, y) \in V(P_k)$,

$$U'_P\Psi'_t(x, y) = (I_k \otimes S_P F_P) \sum_{j \in K_k, u \in l_{x,y}} \alpha'_t(\in'_j, x, y, u)|\in'_j\rangle|x, y, u\rangle = \sum_{j \in K_k, u \in l_{x,y}} \alpha'_t(\in'_j, x, y, u)|\in'_j\rangle(S_P F_P|x, y, u\rangle)$$

On the other hand

$$U_P\Psi_t(x, y) = U_P\Lambda(\psi)\Psi'_t(x, y) = S_{:P} F_P \Lambda(\psi) \sum_{j \in K_k, u \in l_{x,y}} \alpha'_t(\in'_j, x, y, u)|\in'_j\rangle|x, y, u\rangle$$

$$= S_{:P} F_P \sum_{j \in K_k, u \in l_{x,y}} \psi'_j \alpha'_t(\in'_j, x, y, u)|\in'_j\rangle|x, y, u\rangle = \Lambda(\psi) \sum_{j \in K_k, u \in l_{x,y}} \alpha'_t(\in'_j, x, y, u)|\in'_j\rangle(S_P F_P|x, y, u\rangle)$$

$$= \Lambda(\psi)U'_P\Psi'_t(x, y)$$

Since this holds for any $x, y$, then $\Psi_{t+1}(x, y) = \Lambda(\psi)\Psi'_{t+1}(x, y)$, and the proof is finished.

For any $(x, y) \in V(P_k)$, associate the event " $P'_{t,k} = x, P'_{t,k} = y$" by

$P(P'_{t,k} = x, P'_{t,k} = y) = \|\Lambda(\psi)\Psi'_t(x, y)\|^2$. Lemma 2 implies

$P(P'_{t,k} = x, P'_{t,k} = y) = P(P_{t,k} = x, P_{t,k} = y)$ for any $(x, y) \in V(P_k)$. In particular for any initial state of $P_{t,k}$ it is enough to consider the initial state $\sum_{j \in K_k} |\epsilon'_j\rangle |h_j(1,0), Right\rangle$ or

$\sum_{j \in K_k} |\epsilon'_j\rangle |h_j(0,1), Up\rangle$ for $P'_{t,k}$. Now we introduce $(X^*_t, Y^*_t)$ as a reduction of $P'_{t,k}$ on a quarter plane. Here $(X^*_t, Y^*_t)$ is defined on a Hilbert space generated by the following new basis. For all $l \in \{Right, Left, Up, Down\}$ and $x, y \in Z_+$, $|Own_L, 0, 0, \in\rangle = \sum_{j \in K_k} |\epsilon'_j, 0, 0, \epsilon_j\rangle$,

$|Own_D, 0, 0, \in\rangle = \sum_{j \in K_k} |\epsilon'_j, 0, 0, \epsilon_j\rangle$, $|Other_R, 0, 0, \in\rangle = \sum_{j \in K_k} \sum_{k \in K_k \setminus \{j\}} |\epsilon'_k, 0, 0, \epsilon_j\rangle$,

$|Other_U, 0, 0, \in\rangle = \sum_{j \in K_k} \sum_{k \in K_k \setminus \{j\}} |\epsilon'_k, 0, 0, \epsilon_j\rangle$, $|Own_L, x, y, l\rangle = \sum_{j \in K_k} |\epsilon'_j, h_j(x, y), l\rangle$,

$|Own_D, x, y, l\rangle = \sum_{j \in K_k} |\epsilon'_j, h_j(x, y), l\rangle$, $|Other_R, x, y, l\rangle = \sum_{j \in K_k} \sum_{k \in K_k \setminus \{j\}} |\epsilon'_k, h_j(x, y), l\rangle$,

$|Other_U, x, y, l\rangle = \sum_{j \in K_k} \sum_{k \in K_k \setminus \{j\}} |\epsilon'_k, h_j(x, y), l\rangle$. On this basis a step of the time evolution is given by,

for $x, y \in Z_+$,

$|Own_L, 0, 0, \in\rangle \mapsto \tilde{c} a_k^2 |Own_L, 0, 1, Right\rangle + \tilde{c} a_k b_k |Other_R, 0, 1, Right\rangle + \tilde{c} a_k b_k |Own_D, 0, 1, Right\rangle + \tilde{c} b_k^2 |Other_U, 0, 1, Right\rangle$

$|Own_D, 0, 0, \in\rangle \mapsto \tilde{c} a_k b_k (k-1) |Own_L, 0, 1, Right\rangle + \tilde{c} b_k^2 (k-1) |Other_R, 0, 1, Right\rangle - \tilde{c} a_k^2 |Own_D, 0, 1, Right\rangle - \tilde{c} a_k b_k |Other_U, 0, 1, Right\rangle$

$|Other_R, 0, 0, \in\rangle \mapsto \tilde{c} a_k b_k (k-1) |Own_L, 0, 1, Right\rangle - \tilde{c} a_k^2 |Other_R, 0, 1, Right\rangle + \tilde{c} b_k^2 (k-1) |Own_D, 0, 1, Right\rangle - \tilde{c} a_k b_k |Other_U, 0, 1, Right\rangle$

$|Other_U, 0, 0, \in\rangle \mapsto \tilde{c} b_k^2 (k-1)^2 |Own_L, 0, 1, Right\rangle - \tilde{c} a_k b_k (k-1) |Other_R, 0, 1, Right\rangle - \tilde{c} a_k b_k (k-1) |Own_D, 0, 1, Right\rangle + \tilde{c} a_k^2 |Other_U, 0, 1, Right\rangle$

and for each $m \in \{Own_L, Own_D, Other_R, Other_U\}$, we have

$|m, x, y, Left\rangle \mapsto a^2 |m, x-1, y, Left\rangle + ac|m, x+1, y, Right\rangle + ac|m, x, y-1, Down\rangle + c^2 |m, x, y+1, Up\rangle$

$|m, x, y, Right\rangle \mapsto ab|m, x-1, y, Left\rangle + ad|m, x+1, y, Right\rangle + bc|m, x, y-1, Down\rangle + cd|m, x, y+1, Up\rangle$

$$|m,x,y,Down\rangle \mapsto ab|m,x-1,y,Left\rangle + bc|m,x+1,y,Right\rangle + ad|m,x,y-1,Down\rangle +$$
$$cd|m,x,y+1,Up\rangle$$
$$|m,x,y,Up\rangle \mapsto b^2|m,x-1,y,Left\rangle + bd|m,x+1,y,Right\rangle + bd|m,x,y-1,Down\rangle +$$
$$d^2|m,x,y+1,Up\rangle$$

Note that we can write the evolution operator of $(X_t^*, Y_t^*)$ as say $U_H^* = F_H^* S_H^*$ since the subspace generated by this basis is invariant under the operation $U_P'$ and we can write the initial state of $P_{t,k}'$ as $|Own_L, 0, 1, Right\rangle$. Note that $F_H^* = |0,0\rangle\langle 0,0| \otimes \tilde{c} C_H^* + \sum_{x,y \in Z_+} |x,y\rangle\langle x,y| \otimes C_{Z \times Z}$, where $C_{Z \times Z}$ is as defined earlier on,
$$C_H^* = \begin{bmatrix} a_k^2 & a_k b_k (k-1) & a_k b_k (k-1) & b_k^2 (k-1)^2 \\ a_k b_k & -a_k^2 & b_k^2 (k-1) & -a_k b_k (k-1) \\ a_k b_k & (k-1)b_k^2 & -a_k^2 & -a_k b_k (k-1) \\ b_k^2 & -a_k b_k & -a_k b_k & a_k^2 \end{bmatrix}$$
, and $\tilde{c} \in C$ is such that $|\tilde{c}| = 1$. For $m \in \{Own_L, Own_D, Other_R, Other_U\}$ and $l \in \{Right, Left, Up, Down\}$, the shift operator $S_H^*$ is given by $S_H^* |m,0,0,\in\rangle = |m,1,0,Right\rangle + |m,0,1,Up\rangle$,

$$S_H^* |m,1,0,l\rangle = \begin{cases} |m,0,0,\in\rangle, l = Left \\ |m,2,0,Right\rangle, l = Right \end{cases}, \quad S_H^* |m,0,1,l\rangle = \begin{cases} |m,0,0,\in\rangle, l = Down \\ |m,0,2,Up\rangle, l = Up \end{cases},$$

$$S_H^* |m,x,y,l\rangle = \begin{cases} |m,x-1,y,Left\rangle, l = Left \\ |m,x+1,y,Right\rangle, l = Right \\ |m,x,y-1,Down\rangle, l = Down \\ |m,x,y+1,Up\rangle, l = Up \end{cases}.$$

Let $\Psi_t^*(x,y)$ be the state of the quantum walk $(X_t^*, Y_t^*)$. Note that we can write the initial state as $\Psi_1^*(0,1) = |Own_L, 0, 1, Right\rangle$. Now define for $r \in K_k$,

$$\Lambda(\psi) = \left( \psi_r' \{ \langle Own_L | + \langle Own_D | \} + \sum_{j \in K_k \setminus \{r\}} \psi_j' \{ \langle Other_R | + \langle Other_U | \} \right) \otimes I_P. \text{ Let } (X_{t,r}, Y_{t,r}) \text{ be}$$

associated with the event "$X_{t,r} = x, Y_{t,r} = y$", where $P(X_{t,r} = x, Y_{t,r} = y) = \|\Lambda(\psi) \Psi_t^*(x,y)\|^2$. For $t \geq 1$, the probability is determined by $\Psi_t(h_r(x,y))$ in the following way

$$P(X_{t,r} = x, Y_{t,r} = y) = \begin{cases} |\alpha_t(0,0,\in_r)|^2, & x=0; y=0 \\ \|\Psi_t(h_r(x,y))\|^2, & otherwise \end{cases}$$

. The relation between the events

"$P_{t,k} = h_r(x,y)$" and "$X_{t,r} = x, Y_{t,r} = y$" is given by

$$P(P_{t,k} = h_r(x,y)) = \begin{cases} \sum_{j \in K_k} P(X_{t,j} = 0, Y_{t,j} = 0), & x=0; y=0 \\ P(X_{t,r} = x, Y_{t,r} = y), & otherwise \end{cases}$$

. It should be noted that

$$\sum_{j \in K_k} \sum_{x,y \in \{0\} \cup Z_+} P(X_{t,j} = 0, Y_{t,j} = 0) = 1.$$

**VII. The Generating Function of $\Psi_t^*(x,y)$**

Let $m \in \{Own_L, Own_D, Other_R, Other_U\}$. Denote $|m\rangle|0,0,\in\rangle$ as $|m\rangle|0,0,Left\rangle$ and construct $|m\rangle|0,0,Right\rangle$ as the dummy base. Diagrammatically we can express the evolution operator of the walk using weights and arrows, the weights are given by

$$Q_0' = \tilde{Q} = \begin{bmatrix} a_k^2 & a_k b_k(k-1) & a_k b_k(k-1) & b_k^2(k-1)^2 \\ a_k b_k & -a_k^2 & b_k^2(k-1) & -a_k b_k(k-1) \\ a_k b_k & (k-1)b_k^2 & -a_k^2 & -a_k b_k(k-1) \\ b_k^2 & -a_k b_k & -a_k b_k & a_k^2 \end{bmatrix} \otimes \begin{bmatrix} 0 & 0 & 0 & 0 \\ 0 & 0 & 0 & 0 \\ 0 & 0 & 0 & 0 \\ \tilde{c}^2 & 0 & 0 & 0 \end{bmatrix},$$

$$P_L = I_4 \otimes \begin{bmatrix} a^2 & ab & ab & b^2 \\ 0 & 0 & 0 & 0 \\ 0 & 0 & 0 & 0 \\ 0 & 0 & 0 & 0 \end{bmatrix}, P_R = I_4 \otimes \begin{bmatrix} 0 & 0 & 0 & 0 \\ ac & ad & bc & bd \\ 0 & 0 & 0 & 0 \\ 0 & 0 & 0 & 0 \end{bmatrix}, Q_D = I_4 \otimes \begin{bmatrix} 0 & 0 & 0 & 0 \\ 0 & 0 & 0 & 0 \\ ac & bc & ad & bd \\ 0 & 0 & 0 & 0 \end{bmatrix}$$

$$Q_U = I_4 \otimes \begin{bmatrix} 0 & 0 & 0 & 0 \\ 0 & 0 & 0 & 0 \\ 0 & 0 & 0 & 0 \\ c^2 & cd & cd & d^2 \end{bmatrix},$$

$$\Psi_0^*(0,0) = \begin{bmatrix} \dfrac{a_k^2}{\tilde{c}^2} \\ \dfrac{a_k b_k}{\tilde{c}^2} \\ \dfrac{a_k b_k}{\tilde{c}^2} \\ \dfrac{b_k^2}{\tilde{c}^2} \end{bmatrix} \otimes \begin{bmatrix} 1 \\ 0 \\ 0 \\ 0 \end{bmatrix}.$$ Note that $(X_t^*, Y_t^*)$ is a walk starting at $t=1$ and $(1,0)$, in particular

$$\Psi_1^*(1,0) = \begin{bmatrix} 1 \\ 0 \\ 0 \\ 0 \end{bmatrix} \otimes \begin{bmatrix} 0 \\ 0 \\ 0 \\ 1 \end{bmatrix} = \tilde{Q}\Psi_0^*(0,0).$$ We define the generating function for the state by

$$\tilde{\Psi}^*(x,y,z) = \sum_{t=0}^{\infty} \Psi_t^*(x,y) z^t.$$ In order to compute the generating function we first define

$\tilde{\Xi}((0,0) \to (x,y), \tau)$ as the weight of all paths starting from $(0,0)$ ending at $(x,y)$ after $\tau$ steps, and $\Xi((0,0) \to (x,y), \tau)$ as the weight of all paths on another walk defined by $Q_0' = P_R$. We now calculate the generating function for $\Xi((0,0) \to (x,y), \tau)$. Since the first operator could be $P_R$ or $Q_U$, the weights of path form $P_R \cdots P_R$ or $Q_U \cdots Q_U$ or $P_R \cdots Q_U$ or $Q_U \cdots P_R$. So we express $\Xi((0,0) \to (x,y), \tau)$ as a linear combination of $P_R, Q_U, P_R', Q_R'$ giving

$$\Xi((0,0) \to (x,y), \tau) = b^{p_R}((0,0) \to (x,y); \tau) P_R + b^{q_U}((0,0) \to (x,y); \tau) Q_U + b^{p_R'}((0,0) \to (x,y); \tau) P_R' + b^{q_U'}((0,0) \to (x,y); \tau) Q_U' + \delta_0(x,y)\delta(\tau) I_4 \otimes I_4$$

where $P_R' = I_4 \otimes \begin{bmatrix} 0 & 0 & 0 & 0 \\ 0 & 0 & 0 & 0 \\ ac & ad & bc & bd \\ 0 & 0 & 0 & 0 \end{bmatrix}$ and $Q_U' = I_4 \otimes \begin{bmatrix} c^2 & cd & cd & cd \\ 0 & 0 & 0 & 0 \\ 0 & 0 & 0 & 0 \\ 0 & 0 & 0 & 0 \end{bmatrix}$, and for each

$j \in \{p_R, q_U, p_R', q_U'\}$ we define $b^j((0,0) \to (x,y); 0) = 0$. The generating function for $\Xi((0,0) \to (x,y), \tau)$ is defined by

$$\sum_{\tau=0}^{\infty} \Xi((0,0) \to (x,y); \tau) z^\tau = B^{p_R}((0,0) \to (x,y); z) P_R + B^{q_U}((0,0) \to (x,y); z) Q_U + B^{p_R'}((0,0) \to (x,y); z) P_R'$$
$$+ B^{q_U'}((0,0) \to (x,y); z) Q_U' + \delta_0(x,y) I_4 \otimes I_4$$

where for each $j \in \{p_R, q_U, p'_R, q'_U\}$ we have defined

$$B^j((0,0) \to (x, y); z) = \sum_{\tau=0}^{\infty} b^j((0,0) \to (x, y); \tau) z^\tau.$$ Since the left-hand tensor product of

$P_L, P_R, Q_D, Q_U$ is $I_4$, the generating function for $\Xi((0,0) \to (x, y); \tau)$ corresponds to an analogue of the result in [23] on the square lattice. We should remark that the result in Oka et.al is largely determined by equation (24) in Chisaki et.al [5], namely $\lambda(z)$. However this requires solving the characteristic equation associated with a system of difference equations in reference [14] of Oka et.al [23]. Since we can map the result of Oka et.al [23] onto the square lattice, we mainly need to determine the roots of the characteristic equation from the corresponding system of difference equations for the walk on the square lattice. In particular for sufficiently small $z$, we have

$$B^{p_R}((0,0) \to (x, y); z) = \left\{ \frac{d^2 \lambda(z)}{a^2} \right\}^{x+y} \frac{1}{d^2}; x \geq 1, y \geq 0,$$

$$B^{q_U}((0,0) \to (x, y); z) = \left\{ \frac{d^2 \lambda(z)}{a^2} \right\}^{x+y} \frac{1}{d^2}; x \geq 0, y \geq 1,$$

$$B^{p_R, q_U}((0,0) \to (0,0); z) = 0$$

$$B^{p'_R}((0,0) \to (x, y); z) = \left\{ \frac{d^2 \lambda(z)}{a^2} \right\}^{x+y} \frac{\lambda(z) - a^2 z^2}{a^2 c^2 z^2}; x \geq 0, y \geq 0$$

$$B^{q'_U}((0,0) \to (x, y); z) = \left\{ \frac{d^2 \lambda(z)}{a^2} \right\}^{x+y} \frac{\lambda(z) - a^2 z^2}{a^2 c^2 z^2}; x \geq 0, y \geq 0$$

where $\lambda(z) = \dfrac{\Delta z^2 + 1 - \sqrt{\Delta^2 z^4 + 2\Delta(1 - 2|c|^2) z^2 + 1}}{2\Delta \bar{c} z}$. Since $\left|\dfrac{a^2}{d^2}\right| = 1$, we can take $r_0 < 1$ such

that $|\lambda(z)| < 1$ for $|z| < r_0$. Next we calculate the generating function for $\tilde{\Xi}((0,0) \to (0,0), \tau)$. To do so we introduce $\tilde{\Xi}((0,0) \to (0,0), \tau; n)$ as the weight of all paths starting from the origin $n$ times before ending at the origin at time $\tau$. Consider $\tilde{\Xi}((0,0) \to (0,0), \tau; 0)$, for $\tau \geq 2$ we obtain

$$\tilde{\Xi}((0,0) \to (0,0), \tau; 0) = (1 - \delta_2(\tau)) \{P_L b^{p'_R}((0,0) \to (x, y); \tau - 2) P'_R) \tilde{P}_R + P_D b^{q'_U}((0,0) \to (x, y); \tau - 2) Q'_U) \tilde{Q}_U \}$$
$$+ \delta_2(\tau) \{P_L \tilde{P}_R + P_D \tilde{Q}_U\}$$

where $\tilde{P}_R = \begin{bmatrix} a_k & (k-1)b_k \\ b_k & -a_k \end{bmatrix}^{\otimes 2} \otimes \begin{bmatrix} \tilde{c} & 0 \\ 0 & 0 \end{bmatrix}$, and $\tilde{Q}_U$ is defined in a similar way. For $\tau < 2$, we

define

$\tilde{\Xi}((0,0) \to (0,0); \tau, 0) = 0$. So the generating function for $\tilde{\Xi}((0,0) \to (0,0); \tau, 0)$ is given by

$$\sum_{\tau=0}^{\infty} \Xi((0,0) \to (0,0); \tau, 0) z^{\tau} = P_L P_R' \tilde{P}_R B^{p_R'}((0,0) \to (0,0); z) z^2 + P_D Q_U' \tilde{Q}_U B^{q_U'}((0,0) \to (0,0); z) z^2$$
$$+ \left( P_L \tilde{P}_R + P_D \tilde{Q}_U \right) z^2$$

We should remark that the above can be written in simplified form involving few operators via multiplication rules for the $P_L P_R Q_U Q_D P_L' P_R' Q_U' Q_D'$ matrices, an analogue of the multiplication rules for the $PQRS$ matrices in reference [14] of Oka et.al [23]. In our paper we do not consider the simplification, since the size of the matrices makes this time consuming.

Similarly, for $\tau \geq 4$, we have

$$\tilde{\Xi}((0,0) \to (0,0); \tau; 1) = \sum_{\tau_1 + \tau_2 + 4} \begin{pmatrix} (1 - \delta_2(\tau_1)) \{ P_L P_R' \tilde{P}_R b^{p_r'}((0,0) \to (0,0); \tau_1) + P_D Q_U' \tilde{Q}_U b^{q_U'}((0,0) \to (0,0); \tau_1) \} \\ + \delta_2(\tau_1) \{ P_L \tilde{P}_R + P_D \tilde{Q}_U \} \end{pmatrix}$$

$$\begin{pmatrix} (1 - \delta_2(\tau_2)) \{ P_L P_R' \tilde{P}_R b^{p_r'}((0,0) \to (0,0); \tau_2) + P_D Q_U' \tilde{Q}_U b^{q_U'}((0,0) \to (0,0); \tau_2) \} \\ + \delta_2(\tau_2) \{ P_L \tilde{P}_R + P_D \tilde{Q}_U \} \end{pmatrix}$$

and for $\tau < 4$ we define $\tilde{\Xi}((0,0) \to (0,0); \tau; 1) = 0$. Thus the generating function for $\tilde{\Xi}((0,0) \to (0,0); \tau; 1)$ is obtained as

$$\sum_{\tau=0}^{\infty} \tilde{\Xi}((0,0) \to (0,0); \tau; 1) z^{\tau} = \tilde{c}^2 \{ P_L P_R' \tilde{P}_R B^{p_R'}((0,0) \to (0,0); z) z^2 + P_D Q_U' \tilde{Q}_U B^{q_U'}((0,0) \to (0,0); z) z^2 + \left( P_L \tilde{P}_R + P_D \tilde{Q}_U \right) z^2 \}^2$$

Recursively we the following formula for $n \geq 0$,

$$\tilde{B}^{p_R'}((0,0) \to (0,0); z; n) = \left( \frac{1 + (-1)^n}{2} \right) \frac{\{ \tilde{c}^2 \left( P_L P_R' \tilde{P}_R B^{p_R'}((0,0) \to (0,0), z) z^2 \right) \}^{n+1}}{\tilde{c}^2}$$

$$\tilde{B}^{q_U'}((0,0) \to (0,0); z; n) = \left( \frac{1 + (-1)^n}{2} \right) \frac{\{ \tilde{c}^2 \left( P_D Q_U' \tilde{Q}_U B^{q_U'}((0,0) \to (0,0), z) z^2 \right) \}^{n+1}}{\tilde{c}^2}$$

From the two equations immediately above we get the generating function for $\tilde{\Xi}((0,0) \to (0,0); \tau)$ by summing over $n$. We should remark that for $|z| < r_1 = \min(|c^2|, r_0)$ we have

$\left| P_D Q_U' \tilde{Q}_U B^{q_U'}((0,0) \to (0,0), z) z^2 - \Delta z \right| < 1$ and $\left| P_L P_R' \tilde{P}_R B^{p_R'}((0,0) \to (0,0), z) z^2 - \Delta z \right| < 1$. So for $z$ such that $|z| < r_1$,

$$\sum_{\tau=0}^{\infty} \Xi((0,0) \to (0,0); \tau) z^{\tau} = B^{\tilde{p}_R'}((0,0) \to (0,0); z) \tilde{P}_R + B^{\tilde{q}_U'}((0,0) \to (0,0), z) \tilde{Q}_U + I_4 \otimes I_4, \text{ where}$$

$B^{\tilde{p}_r'}((0,0) \to (0,0); z) = \sum_{n=0}^{\infty} B^{p_r'}((0,0) \to (0,0); z; n)$ and $B^{\tilde{q}_u'}((0,0) \to (0,0); z)$ is defined in a similar way. Note that $\tilde{\Xi}((0,0) \to (x, y); \tau)$ is written by $\Xi((0,0) \to (x, y); \tau)$ and

$\tilde{\Xi}((0,0) \to (x, y); \tau)$ as

$$\tilde{\Xi}((0,0) \to (x, y); \tau) = \sum_{\tau_1 + \tau_2 + 1} \Xi((0,0) \to (x-1, y); \tau_2) \tilde{P}_R \tilde{\Xi}((0,0) \to (0,0); \tau_1) +$$

$$\Xi((0,0) \to (x, y-1); \tau_2) \tilde{P}_L \tilde{\Xi}((0,0) \to (0,0); \tau_1) + \delta_0(x, y) I_4 \otimes I_4$$

From the generating function for $\Xi((0,0) \to (x, y); \tau)$ and $\tilde{\Xi}((0,0) \to (0,0); \tau)$ we can compute the generating function for $\tilde{\Xi}((0,0) \to (x, y); \tau)$ as follows: Let $j \in \{p_R, q_U, p'_R, q'_U\}$, and let the corresponding operator be given by $J \in \{P_R, Q_U, P'_R, Q'_U\}$. For example for $j = p_R$, the corresponding operator is $J = P_R$, so that the summation over $j, J$ below is understood to be taken in only one way, then we can write

$$\sum_{\tau=0}^{\infty} \Xi((0,0) \to (x, y); \tau) z^{\tau} = \left( \sum_{j, J} B^j((0,0) \to (x-1, y); z)) J + B^j((0,0) \to (x, y-1); z)) J + \delta_1(x, y) I_4 \otimes I_4 \right) \tilde{Q} z$$

So we obtain the generating function as $\tilde{\Psi}^*(x, y; z) = \sum_{\tau=0}^{\infty} \tilde{\Xi}((0,0) \to (x, y); \tau) z^{\tau} \Psi_0^*(x, y)$

**VIII. Towards a Localization Criterion**

We compute the limit state of $(X_t^*, Y_t^*)$ from the generating function for the state which we defined very early on in the previous section as $\tilde{\Psi}^*(x, y, z) = \sum_{t=0}^{\infty} \Psi_t^*(x, y) z^t$. Note that this generating function can be written as $\tilde{\Psi}^*(x, y, z) = \sum_{\substack{m \in \{Own_L, Own_D, Other_R, Other_U\} \\ l \in \{Left, Right, Up, Down\}}} |m, x, y, l; z\rangle |m, x, y, l\rangle$. Now let

$$\lambda(z) = \frac{\Delta z^2 + 1 - \sqrt{\Delta^2 z^4 + 2\Delta(1 - 2|c|^2) z^2 + 1}}{2 \Delta \bar{c} z}, \quad \mu(z) = \frac{d^2 \lambda(z) - \Delta^2 z}{c^2},$$

$$v(z) = (1 + \Delta z^2)^2 - 4\Delta |a^2|^2 z^2, \quad \Delta_{\pm}(z) = \frac{2c^2}{\tilde{c}^2} \pm 1 \mp \Delta z^2, \quad w_{\pm}^2 = \mp \frac{c^2(\tilde{c}^2 \pm c^2)}{\tilde{c}^2 \Delta(\tilde{c}^2 |a^2|^2 - \tilde{c}^2 \mp c^2)},$$

$\Delta = ad - bc$, then there exists $0 < r_1 < 1$, so that for any $z$ with $|z| < r_1$, we have the following, assuming a one-to-one correspondence between $m \in \{Own_L, Own_D, Other_R, Other_U\}$ and $s \in \{a_k^2, a_k b_k, a_k b_k, b_k^2\}$,

$$\tilde{\alpha}^*(m, x, y, Left; z) = \begin{cases} \dfrac{d^2}{a^2 c^2} (\lambda(z) - a^2 z)(\tilde{c}^2 s \mu(z) + 1) \phi(x, y; z), & x > 0, y > 0 \\ -\mu(z)(\tilde{c}^2 s \mu(z) + 1) \phi(x, y; z), & x = y = 0 \end{cases}$$

$$\tilde{\alpha}^*(m,x,y,Right;z) = \begin{cases} -z(\tilde{c}^2 s\mu(z)+1)\phi(x,y;z), x>0, y>0 \\ 0, x=y=0 \end{cases}$$

$$\tilde{\alpha}^*(m,x,y,Down;z) = \begin{cases} \dfrac{c^2}{b^2 d^2}(\lambda(z)-b^2 z)(\tilde{c}^2 s\mu(z)+1)\phi(x,y;z), x>0, y>0 \\ \mu(z)(\tilde{c}^2 s\mu(z)+1)\phi(x,y;z), x=y=0 \end{cases}$$

$$\tilde{\alpha}^*(m,x,y,Up;z) = \begin{cases} z(\tilde{c}^2 s\mu(z)+1)\phi(x,y;z), x>0, y>0 \\ 0, x=y=0 \end{cases}$$

$$\phi(x,y;z) = \dfrac{\left(\dfrac{d^2\lambda(z)}{a^2}\right)^{x+y-2}\tilde{c}^6 w_+^2 w_-^2\left(\eta_+(z)+\sqrt{v(z)}\right)\left(\eta_-(z)-\sqrt{v(z)}\right)}{4(\tilde{c}^4-c^4)(z^2-w_+^2)(z^2-w_-^2)}$$. From Cauchy's Theorem for

$0<r<r_1<1$, $\Psi_t^*(x,y) = \dfrac{1}{2\pi i}\oint_{|z|=r}\tilde{\Psi}^*(x,y;z)\dfrac{dz}{z^{t+1}}$. Let $g\in\{w_+,-w_+,w_-,-w_-\}$, then as $t\to\infty$,

$$\Psi_t^*(x,y) \sim -\sum_g \operatorname{Res}(\tilde{\Psi}^*(x,y;z),g)g^{-(t+1)},$$ where $\operatorname{Res}(f(z),w)$ is the residue for $f(z)$ for

$z=w$. Now taking the residues of the generating functions, we have the limit states as follows:

Let $\Phi_\pm(x,y,t) = \left(\dfrac{\Delta w_\pm^2 \mp \dfrac{c^2}{\tilde{c}^2}}{a^2 w_\pm}\right)^{x+y-2} \dfrac{w_+^2 w_-^2 \tilde{c}^{10}|c^2|\cos\theta\pm|c^2|}{w_\pm^{t+1}\Delta(1\pm|c^2|e^{-i\phi})^2}$, where $I_+(c^2,\phi) = I_{(-|c^2|,1]}(\cos\phi)$

and $I_-(c^2,\phi) = I_{[-1,|c^2|)}(\cos\phi)$, and assume that there is a 1-1 correspondence between the sets

$\{Own_L, Own_D\} \leftrightarrow \{a_k^2, a_k b_k\} = h$ and $\{Other_R, Other_U\} \leftrightarrow \{a_k b_k, b_k^2\} = q$.

If $m \in \{Own_L, Own_D\}$ and $l \in \{Left, Down\}$, then

$$\alpha_t^*(m,x,y,l) \sim \left(\dfrac{1+(-1)^{x+y+2t}}{2}\right)\begin{cases} I_+(c^2,\phi)\dfrac{\Delta w_+^2|c^2|^2 - \dfrac{c^2}{\tilde{c}^2}}{a^2 c^2 w_+^2}(1-h)\Phi_+(x,y,t) - I_-(c^2,\phi)\dfrac{\Delta w_-^2|c^2|^2 - \dfrac{c^2}{\tilde{c}^2}}{a^2 c^2 w_-^2}(1+h)\Phi_-(x,y,t); x,y \geq 0 \\ \\ I(c^2,\phi)\dfrac{1}{\tilde{c}^2 w_+}(1-h)\Phi_+(x,y,t) - I_-(c^2,\phi)\dfrac{1}{\tilde{c}^2 w_-}(1+h)\Phi_-(x,y,t); x=y=0 \end{cases}$$

If $m \in \{Other_R, Other_U\}$ and $l \in \{Left, Down\}$, then

$$\alpha_t^*(m,x,y,l) \sim \left(\frac{1+(-1)^{x+y+2t}}{2}\right) \begin{cases} -I_+(c^2,\phi)\dfrac{\Delta w_+^2|c^2|^2 - \dfrac{c^2}{\tilde{c}^2}}{a^2 c^2 w_+^2}(q)\Phi_+(x,y,t) - I_-(c^2,\phi)\dfrac{\Delta w_-^2|c^2|^2 - \dfrac{c^2}{\tilde{c}^2}}{a^2 c^2 w_-^2}(q)\Phi_-(x,y,t); x,y \geq 0 \\ \\ -I_+(c^2,\phi)\dfrac{1}{\tilde{c}^2 w_+}(q)\Phi_+(x,y,t) - I_-(c^2,\phi)\dfrac{1}{\tilde{c}^2 w_-}(q)\Phi_-(x,y,t); x = y = 0 \end{cases}$$

If $m \in \{Own_L, Own_D\}$ and $l \in \{Right, Up\}$, then

$$\alpha_t^*(m,x,y,l) = \left(\frac{1+(-1)^{x+y+2t}}{2}\right) \begin{cases} I_+(c^2,\phi)(1-h)\Phi_+(x,y,t) - I_-(c^2,\phi)(1+h)\Phi_-(x,y,t); x,y \geq 0 \\ \\ 0; x = y = 0 \end{cases}$$

If $m \in \{Other_R, Other_U\}$ and $l \in \{Right, Up\}$, then

$$\alpha_t^*(m,x,y,l) = \left(\frac{1+(-1)^{x+y+2t}}{2}\right) \begin{cases} -I_+(c^2,\phi)(q)\Phi_+(x,y,t) - I_-(c^2,\phi)(q)\Phi_-(x,y,t); x,y \geq 0 \\ \\ 0; x = y = 0 \end{cases}$$

After some calculations with $P(X_{t,r} = x, Y_{t,r} = y) = \|\Lambda(\psi)\Psi_t^*(x,y)\|^2$ and the above limit states we arrive at the localization criterion.

Let $\phi = \arg\left(\dfrac{c^2}{\tilde{c}^2}\right)$, $K_\pm = \left|1 + |c^2|e^{-i\phi}\right|^2$, $K_\times = \left(1 - |c^2|e^{i\phi}\right)\left(1 + |c^2|e^{-i\phi}\right)$. Recall that $\Psi_t(x,y)$ is the state of the quantum walk $P_{t,k}$ at time $t$ and position $(x,y)$. For $r \in K_k$, define

$$\psi = \Psi_0(0,0) = \sum_{j \in K_K} \psi_j |0,0,\in_j\rangle, \quad \psi_r'(\psi) = \sum_{j \in K_K} \tilde{c}\left(\frac{2}{k} - \delta_r(j)\right)\psi_j, \quad \theta_1^r(\psi) = |\psi_r'(\psi)|^2,$$

$$\theta_2^r(\psi) = \sum_{j \in K_r \setminus \{r\}} \overline{\psi_r'(\psi)}\psi_j'(\psi), \quad \theta_3^r(\psi) = \sum_{\substack{j,k \in K_r \setminus \{r\} \\ j \neq k}} \overline{\psi_k'(\psi)}\psi_j'(\psi) + \overline{\psi_j'(\psi)}\psi_k'(\psi) + \sum_{j \in K_k \setminus \{r\}} |\psi_j'(\psi)|^2, \text{ where}$$

$\delta_r(j)$ is the indicator function at $r$. The localization criterion is contained in the following theorem.

**Theorem 1(Localization):** For $k \geq 1$, $x, y \in Z_+ \cup \{0\}$, $r \in K_k$,

$$P(X_{t,r} = x, Y_{t,r} = y) \sim \frac{1+(-1)^{x+y+2t}}{2}\left\{I_{[-1,|c^2|]}(\cos\phi)L_m(x,y) + I_{(-|c^2|,1]}(\cos\phi)L_p^r(x,y) + I_{(-|c^2|,|c^2|)}(\cos\phi)L_c^r(x,y,t)\right\}$$

where $f(t) \sim g(t)$ means $\lim_{t\to\infty} \frac{f(t)}{g(t)} = 1$, and $L_m(x,y) = \Gamma_-(x,y)\left|\sum_{j=0}^{k-1}\psi_j\right|^2$,

$$L_p^r(x,y) = \Gamma_+(x,y)\left|\sum_{j=0}^{k-1}(\psi_j - \psi_r)\right|^2,$$

$$L_c^r(x,y,t) = 2\sum_h -(1-h^2)\theta_1^r(\psi)\text{Re}(\Gamma_\times(x,y,t)) + 2\sum_{h,q} q(1+h)\text{Re}(\theta_2^r(\psi)\Gamma_\times(x,y,t)) - 2\sum_{h,q} q(1-h)\text{Re}(\theta_2^r(\psi)\overline{\Gamma_\times(x,y,t)}) + 2\sum_q q^2\theta_3^r(\psi)\text{Re}(\Gamma_\times(x,y,t))$$

$$\Gamma_\pm(x,y) = \frac{\sum_q q^2|c|^4(\cos\phi \pm |c^2|)^2}{K_\pm^2}\left\{\delta_0(x,y) + (1-\delta_0(x,y))\left(\frac{|a|^4}{K_\pm}\right)^{x+y-2}\left(1+\frac{|a|^4}{K_\pm}\right)\right\}$$

$$\Gamma_\pm(x,y,t) = \left(\sqrt{\frac{K_\times}{K_\times}}\right)\frac{|c|^4(\cos^2\phi - |c|^4)}{K_\times^2}\left\{-\delta_0(x,y)\sqrt{\frac{K_\times}{K_\times}} + (1-\delta_0(x,y))\left(\frac{|a|^4}{\sqrt{K_+K_-}}\right)^{x+y-2}\left(1-\frac{|a|^4}{K_\times}\right)\right\}$$

We should remark that $L_c^r(x,y,t)$ plays a similar role to the $\delta$ – measure corresponding to localization in the following papers [6,24] for example. In particular $L_c^r(x,y,t)$ is an oscillatory term corresponding to localization.

## IX. Towards a Weak Limit Theorem

We calculate the Fourier transform of the generating function as

$$\hat{\tilde{\Psi}}^*(s_x, s_y; z) = \sum_{x,y} \tilde{\Psi}^*(x,y;z)e^{is_x x + is_y y}$$ from the following relations

$$\tilde{\alpha}^*(m,x,y,Left;z) = \begin{cases} \frac{d^2}{a^2c^2}(\lambda(z) - a^2z)(\tilde{c}^2 s\mu(z) + 1)\phi(x,y;z), x > 0, y > 0 \\ -\mu(z)(\tilde{c}^2 s\mu(z) + 1)\phi(x,y;z), x = y = 0 \end{cases}$$

$$\tilde{\alpha}^*(m,x,y,Right;z) = \begin{cases} -z(\tilde{c}^2 s\mu(z) + 1)\phi(x,y;z), x > 0, y > 0 \\ 0, x = y = 0 \end{cases}$$

$$\tilde{\alpha}^*(m,x,y,Down;z) = \begin{cases} \frac{c^2}{b^2d^2}(\lambda(z) - b^2z)(\tilde{c}^2 s\mu(z) + 1)\phi(x,y;z), x > 0, y > 0 \\ \mu(z)(\tilde{c}^2 s\mu(z) + 1)\phi(x,y;z), x = y = 0 \end{cases}$$

$$\tilde{\alpha}^*(m,x,y,Up;z) = \begin{cases} z(\tilde{c}^2 s\mu(z)+1)\phi(x,y;z), x>0, y>0 \\ 0, x=y=0 \end{cases}$$, then we obtain the characteristic

function from the following relation

$$E\{e^{i\xi X_{t,r}+i\zeta Y_{t,r}}\} = \sum_{x,y \in Z} \langle \Lambda_r(\psi)\Psi_t^*(x,y), \Lambda_r(\psi)\Psi_t^*(x,y) \rangle e^{i\xi x+i\zeta y}$$

$$= \sum_{x,y,x',y' \in Z} \langle \Lambda_r(\psi)\Psi_t^*(x,y), \Lambda_r(\psi)\Psi_t^*(x',y') \rangle e^{i\xi x+i\zeta y} \int_0^{2\pi}\int_0^{2\pi} e^{ik_x(x-x')+ik_y(y-y')} \frac{dk_x}{2\pi}\frac{dk_y}{2\pi}$$

$$= \int_0^{2\pi}\int_0^{2\pi} \left\{ \sum_{x,y,x',y' \in Z} \langle \Lambda_r(\psi)\Psi_t^*(x,y), \Lambda_r(\psi)\Psi_t^*(x',y') \rangle e^{ik_x(x-x')+ik_y(y-y')} e^{i\xi x+i\zeta y} \right\} \frac{dk_x}{2\pi}\frac{dk_y}{2\pi}$$

$$= \int_0^{2\pi}\int_0^{2\pi} \langle \Lambda_r(\psi)\widehat{\Psi}_t^*(s_x,s_y), \Lambda_r(\psi)\widehat{\Psi}_t^*(s_x+\xi,s_y+\zeta) \rangle \frac{ds_x}{2\pi}\frac{ds_y}{2\pi}$$

where $\langle \vec{u},\vec{v} \rangle$ is the inner product of the vectors $\vec{u}, \vec{v}$. Now we write

$$\hat{\tilde{\Psi}}^*(s_x,s_y;z) = \sum_{\substack{m \in \{Own_L, Own_D, Other_R, Other_U\} \\ l \in \{Left, Right, Down, Up\}}} \hat{\tilde{\alpha}}^*(m,l,s_x,s_y;z)|m,l\rangle$$, then it follows from the relations

$$\tilde{\alpha}^*(m,x,y,Left;z) = \begin{cases} \dfrac{d^2}{a^2c^2}(\lambda(z)-a^2z)(\tilde{c}^2 s\mu(z)+1)\phi(x,y;z), x>0, y>0 \\ -\mu(z)(\tilde{c}^2 s\mu(z)+1)\phi(x,y;z), x=y=0 \end{cases}$$

$$\tilde{\alpha}^*(m,x,y,Right;z) = \begin{cases} -z(\tilde{c}^2 s\mu(z)+1)\phi(x,y;z), x>0, y>0 \\ 0, x=y=0 \end{cases}$$

$$\tilde{\alpha}^*(m,x,y,Down;z) = \begin{cases} \dfrac{c^2}{b^2d^2}(\lambda(z)-b^2z)(\tilde{c}^2 s\mu(z)+1)\phi(x,y;z), x>0, y>0 \\ \mu(z)(\tilde{c}^2 s\mu(z)+1)\phi(x,y;z), x=y=0 \end{cases}$$

$$\tilde{\alpha}^*(m,x,y,Up;z) = \begin{cases} z(\tilde{c}^2 s\mu(z)+1)\phi(x,y;z), x>0, y>0 \\ 0, x=y=0 \end{cases}$$

that we have the following,

$$\hat{\tilde{\alpha}}^*(m,Left,s_x,s_y;z) = \left(-\mu(z)+\frac{d^2}{a^2c^2}(\lambda(z)-a^2z)\right)(\tilde{c}^2 s\mu(z)+1)\Phi_1(s_x,s_y;z)$$

$$\hat{\tilde{\alpha}}^*(m,Right,s_x,s_y;z) = z(1+\tilde{c}^2\mu(z))\Phi_1(s_x,s_y;z)\Phi_2(s_x,s_y;z)$$

$$\hat{\tilde{\alpha}}^*(m, Down, s_x, s_y; z) = \left(\mu(z) + \frac{c^2}{b^2 d^2}(\lambda(z) - bz)\right)(\tilde{c}^2 s \mu(z) + 1)\Phi_1(s_x, s_y; z)$$

$$\hat{\tilde{\alpha}}^*(m, Up, s_x, s_y; z) = z(1 + \tilde{c}^2 \mu(z))\Phi_1(s_x, s_y; z)\Phi_2(s_x, s_y; z) \text{, where}$$

$$\Phi_1(s_x, s_y; z) = \frac{\tilde{c}^6 w_+^2 w_-^2 (\eta_+(z) + \sqrt{v(z)})(\eta_-(z) - \sqrt{v(z)})}{4(\tilde{c}^4 - c^4)(z^2 - w_+^2)(z^2 - w_-^2)},$$

$$\Phi_2(s_x, s_y; z) = \frac{e^{ik_x + ik_y}(\gamma(s_x, s_y; z) - \sqrt{v(z)})}{4\Delta(z - v_+(s))(z - v_-(s))}, \quad \gamma(s_x, s_y; z) = 4a^2 e^{-is_x - is_y} z - 1 - \Delta z^2,$$

$$v_\pm(s) = \frac{a^2 e^{-is_x - is_y} + \bar{a}^2 \Delta e^{is_x + is_y} \pm \sqrt{(a^2 e^{-is_x - is_y} + \bar{a}^2 e^{is_x + is_y})^2 - 4\Delta}}{2\Delta}$$

Next we get the Fourier transform of the state, $\hat{\Psi}_t^*(s_x, s_y)$. Since $\left\|\hat{\tilde{\Psi}}_*(s_x, s_y; z)\right\|^2$ is finite for

$0 < z < r_1$, we can write $\hat{\tilde{\Psi}}^*(s_x, s_y; z) = \sum_{t \geq 0} \hat{\Psi}_t^*(s_x, s_y) z^t$. In particular for $0 < z < r_1$,

$\hat{\Psi}_t^*(s_x, s_y) = \frac{1}{2\pi i} \oint_{|z|=r} \hat{\tilde{\Psi}}(s_x, s_y; z) \frac{dz}{z^{t+1}}$, thus we have the following relation for $\hat{\Psi}_t^*(s_x, s_y)$: Put

$g \in \{w_+, w_-, -w_+, -w_-\}$ and $r \in \{v_+, v_-\}$, then

$\hat{\Psi}_t^*(s_x, s_y) \sim -\sum_g \text{Re} s\left(\hat{\tilde{\Psi}}(s_x, s_y; z), g\right) g^{-(t+1)} - \sum_r \text{Re} s\left(\hat{\tilde{\Psi}}(s_x, s_y; z), r\right) r^{-(t+1)}$. Finally using the

expression immediately above and the relation for $E\{e^{i\xi X_{t,r} + i\zeta Y_{t,r}}\}$, we obtain the following equations

$$\int_0^{2\pi}\int_0^{2\pi} \sum_{g \in \{w_+, -w_+\}} \left\|\Lambda_r(\psi) \text{Re} s\left(\hat{\tilde{\Psi}}(s_x, s_y), g\right)\right\|^2 \frac{ds_x}{2\pi} \frac{ds_y}{2\pi} = I_{(-|c^2|,1]}(\cos \phi) \frac{\sum_q q^2 |c|^4 (\cos \phi + |c^2|)}{2K_+} \left|\sum_{j=0}^{k-1}(\psi_j - \psi_r)\right|^2 \equiv C_p^r$$

$$\int_0^{2\pi}\int_0^{2\pi} \sum_{g \in \{w_-, -w_-\}} \left\|\Lambda_r(\psi) \text{Re} s\left(\hat{\tilde{\Psi}}(s_x, s_y), g\right)\right\|^2 \frac{ds_x}{2\pi} \frac{ds_y}{2\pi} = I_{[-1,|c^2|)}(\cos \phi) \frac{\sum_q q^2 |c|^4 (\cos \phi + |c^2|)}{2K_+} \left|\sum_{j=0}^{k-1}\psi_j\right|^2 \equiv C_m$$

and if $g \in \{w_+, w_-\}$ we obtain

$$\int_0^{2\pi}\int_0^{2\pi} \left(\left\langle \Lambda_r(\psi) \text{Re} s\left(\hat{\tilde{\Psi}}(s_x, s_y), g\right), \Lambda_r(\psi) \text{Re} s\left(\hat{\tilde{\Psi}}(s_x, s_y), -g\right)\right\rangle + \left\langle \Lambda_r(\psi) \text{Re} s\left(\hat{\tilde{\Psi}}(s_x, s_y), -g\right), \Lambda_r(\psi) \text{Re} s\left(\hat{\tilde{\Psi}}(s_x, s_y), g\right)\right\rangle\right) \frac{ds_x}{2\pi} \frac{ds_y}{2\pi} = 0$$

From the relation involving $E\{e^{i\xi X_{t,r} + i\zeta Y_{t,r}}\}$ and the Riemann-Lebesgue Lemma we have

$$\lim_{t\to\infty} E\left\{e^{\frac{i\xi X_{t,r}}{t}+\frac{i\zeta Y_{t,r}}{t}}\right\} = C_p^r + C_m + \int_0^{2\pi}\int_0^{2\pi} e^{-i\xi h(s_x,s_y)-i\zeta h(s_x,s_y)} p(s_x,s_y)\frac{ds_x}{2\pi}\frac{ds_y}{2\pi} + \int_0^{2\pi}\int_0^{2\pi} e^{i\xi h(s_x,s_y)+i\zeta h(s_x,s_y)} q(s_x,s_y)\frac{ds_x}{2\pi}\frac{ds_y}{2\pi}$$

where $p(s_x,s_y) = \left\|\Lambda_r(\psi)\operatorname{Re} s\left(\hat{\tilde{\Psi}}(s_x,s_y),v_+\right)\right\|^2$ and $q(s_x,s_y) = \left\|\Lambda_r(\psi)\operatorname{Re} s\left(\hat{\tilde{\Psi}}(s_x,s_y),v_-\right)\right\|^2$,

and $h(s_x,s_y) = h(s_x)\otimes h(s_y)$, where $h(s_x) = \dfrac{|a|\sin\gamma(s_x)}{\sqrt{1-|a|^2\cos^2\gamma(s_x)}}$ and $h(s_y)$ is defined in a similar way. Choosing suitable contours $C_+$ and $C_-$ we can write the last two terms on the right hand side of $\lim_{t\to\infty} E\left\{e^{\frac{i\xi X_{t,r}}{t}+\frac{i\zeta Y_{t,r}}{t}}\right\}$ as

$$\int_0^{2\pi}\int_0^{2\pi} e^{-i\xi h(s_x,s_y)-i\zeta h(s_x,s_y)} p(s_x,s_y)\frac{ds_x}{2\pi}\frac{ds_y}{2\pi} + \int_0^{2\pi}\int_0^{2\pi} e^{i\xi h(s_x,s_y)+i\zeta h(s_x,s_y)} q(s_x,s_y)\frac{ds_x}{2\pi}\frac{ds_y}{2\pi}$$

$$= \iint_{C_+} e^{-i\xi h(s_x,s_y)-i\zeta h(s_x,s_y)} \{p(-s_x,s_y) + p(s_x,-s_y) + p(-s_x,-s_y) + q(s_x,s_y)\}\frac{ds_x}{2\pi}\frac{ds_y}{2\pi}$$

$$+ \iint_{C_-} e^{-i\xi h(s_x,s_y)-i\zeta h(s_x,s_y)} \{p(-s_x,s_y) + p(s_x,-s_y) + p(-s_x,-s_y) + q(s_x,s_y)\}\frac{ds_x}{2\pi}\frac{ds_y}{2\pi}$$

$$= \int_0^\infty\int_0^\infty e^{-i\xi h(s_x,s_y)-i\zeta h(s_x,s_y)} w(x,y) f_H(x,y)\, dx\, dy$$

where $f_H(x,y) = f_H(x)\otimes f_H(y)$, $f_H(x) = \dfrac{I_{[0,|a|)}(x)\sqrt{1-|a|^2}}{\pi(1-x^2)\sqrt{|a|^2-x^2}}$ and $f_H(y)$ is defined in a similar way, $x = h(s_x)$, $y = h(s_y)$. After some computations with $w(x,y)$ from $p(s_x,s_y)$ and $q(s_x,s_y)$ we get the weak limit theorem as follows.

**Theorem 2(Weak Convergence):** For $k \geq 1$, $r \in K_k$, as $t\to\infty$ we have

$$P\left(u \leq \frac{X_{t,r}}{t} \leq v, u' \leq \frac{Y_{t,r}}{t} \leq v'\right) \to \int_{u'}^{v'}\int_u^v \rho_w^r(x,y)\,dx\,dy\,,\text{ where }$$

$$\rho_w^r(x,y) = \left\{I_{[-1,|c^2|)}(\cos\phi)C_m + I_{(-|c^2|,1]}(\cos\phi)C_p^r\right\}\delta_0(x,y) + C_d^r(x,y) f_H(x,y)\,,\ C_m,C_p^r,$$

and $f_H(x,y)$ are as defined earlier on, and $C_d^r(x,y) = C_d^r(x)\otimes C_d^r(y)$, where

$$C_d^r(x) = \frac{\Gamma_1(x)\theta_1^r(\psi) + 2\operatorname{Re}(\Gamma_2(x)\theta_2^r(\psi)) + \Gamma_3(x)\theta_3^r(\psi)}{(K_+ - (1-x^2)\sin^2\phi)(K_- - (1-x^2)\sin^2\phi)}, \text{ where}$$

$$\Gamma_1(x) = 4a_k|c|(|a|^2 - x^2)\cos\phi\sin^2\phi + (a_k^2 + 2a_k|c|\cos\phi + 1)(1 + |c|^2 - 2|c|^2\cos^2\phi - \sin^2\phi(1-x^2))$$

$$\Gamma_2(x) = -2b_k|c|(|a|^2 - x^2)e^{i\phi}\cos\phi\sin\phi + b_k(a_k + |c|e^{i\theta})(1 + |c|^2 - 2|c|^2\cos^2\phi - \sin^2\phi(1-x^2))$$

$$\Gamma_3(x) = b_k^2(1 + |c|^2 - 2|c|^2\cos^2\phi - (1-x^2)\sin^2\phi), \text{ and } C_d^r(y) \text{ is defined similarly.}$$

## X. Concluding Remarks

In this paper we have obtained an explicit expression for the limit probability of $P_{t,k}$, a result corresponding to localization- there exists a site of the graph $v$ such that $\limsup_{t\to\infty} P(P_{t,k} = v) > 0$.

We have also obtained the weak convergence of $P_{t,k}$. Moreover, the limit measure has a density function $f_H(x,y) = f_H(x) \otimes f_H(y)$, where $f_H(x)$ and $f_H(y)$ are both half-line versions of a typical function in the weak convergence of the quantum walk [23-29]. The function is defined by $f_H(x) = \dfrac{I_{[0,|a|)}(x)\sqrt{1-|a|^2}}{\pi(1-x^2)\sqrt{|a|^2 - x^2}}$, where $I_A(x)$ is the indicator function of a set $A$, and $a$ is an element of the evolution operator of the quantum walk on the graph with joined half-lines.